\documentclass[prb,twocolumn,epsfig,showpacs]{revtex4}
\usepackage{graphicx}
\usepackage{epsfig}

\begin{document}

\title{Model for a Macroscopically Disordered Conductor with an
Exactly Linear High-Field Magnetoresistance}
\author{V. Guttal}
\email{vishw@mps.ohio-state.edu}
\author{D. Stroud}
\email{stroud@mps.ohio-state.edu}
\address{Department of Physics, The Ohio State University, Columbus,
Ohio 43210-1106}


\begin{abstract}

We calculate the effective resistivity of a macroscopically
disordered two-dimensional conductor consisting of two components in
a perpendicular magnetic field. When the two components have equal
area fractions, we use a duality theorem to show that the
magnetoresistance is nonsaturating and at high fields varies exactly
linearly with magnetic field. At other compositions, an
effective-medium calculation leads to a saturating
magnetoresistance.  We briefly discuss possible connections between
these results and magnetoresistance measurements on heavily
disordered chalcogenide semiconductors.

\end{abstract}

\maketitle


\newpage

The resistivity of most homogeneous materials (metals or
semiconductors) increases quadratically with magnetic field $H$ at
low fields, and generally saturates at sufficiently large
$H$\cite{am}. Exceptions may occur for materials with Fermi surfaces
allowing open orbits, or for compensated homogeneous semiconductors,
where the resistivity may increase without saturation, usually
proportional to $H^2$\cite{ziman,am}. Under some special conditions,
the magnetoresistance can be linear in magnetic
field\cite{abrikosov00}.

Recently, a remarkably large transverse magnetoresistance (TMR) has
been observed in the doped silver chalcogenides Ag$_{2+\delta}$Se
and Ag$_{2+\delta}$Te\cite{xu,husmann}.  In these materials, over
the temperature range from 4 to 300 K, the resistivity increases
approximately linearly with $H$ up to fields, applied perpendicular
to the direction of current flow,
as high as 60 T. 
Moreover, the TMR is especially large and most clearly linear at
pressures where the Hall resistivity changes sign\cite{lee}.
Because of this linearity, these materials may be useful as magnetic
field sensors even at megagauss fields.

But beyond the possible applications, the origin of the effect
remains mysterious.  According to conventional theories, such narrow
gap semiconductors should have a saturating TMR. Furthermore, since
these materials contain no magnetic moments, a spin-mediated
mechanism seems unlikely.

There are presently two proposed explanations for this quasilinear
TMR. The first is a quantum theory of magnetoresistance
(MR)\cite{abrik}.  The second proposed mechanism\cite{parish} is
that this nonsaturating TMR arises from macroscopic sample
inhomogeneities. Such inhomogeneities could produce large spatial
fluctuations in the conductivity tensor and hence a large TMR,
especially at large $H$.  This explanation seems plausible because
the chalcogenides probably have a granular microstructure
\cite{lee}, and hence a spatially varying conductivity.

The effective conductivity of media, with a spatially varying
conductivity ${\bf \sigma(x)}$, has been studied since the time of
Maxwell, but a relatively few studies have considered the
magnetoresistance\cite{herring,bruls,dreizin,mendel,sb,bergstrel,stroud76,stroud79,balagurov,sarychev}.
For a three-dimensional medium, the TMR of an isotropic metal does
indeed vary linearly in H, when a small volume fraction $p \ll 1$ of
inclusions of a different carrier density is added \cite{stroud76}.
But the TMR generally does not remain strictly linear at higher
concentrations of p. If the inclusions are strictly insulating, then
the TMR does remain asymptotically linear if the TMR is computed
within the effective-medium approximation\cite{bs}, but its exact
behavior is not known even in this case. Recent experiments on
homogeneous semiconductors containing a gold
inhomogeneity\cite{solin} show a hugely enhanced but not strictly
linear room-temperature geometrical TMR (i.\ e.\ arising from
inhomogeneities); this so-called extraordinary magnetoresistance has
been successfully modeled, using finite-element
techniques\cite{moussa}.

The model of Ref. \cite{parish} also assumes a film with a spatially
varying conductivity.  The inhomogeneities are described by an
impedance network; the tensor nature of the magnetoconductivity is
included by making each network element a four-terminal impedance.
Their numerical solution suggests that, for the network to have a
non-saturating TMR one needs (i) carriers of two different signs,
and (ii) a suitably defined average mobility $\langle \mu \rangle
\sim 0$. When solved numerically and averaged over many disorder
realizations, their model does indeed give a nonsaturating,
approximately linear TMR over a broad field range. Obviously, it
would be useful to have {\em exact} analytical statements to compare
with these numerical results.

In this Rapid Communication, we present an idealized model of a
disordered semiconducting film in two dimensions.  The model assumes
a macroscopically inhomogeneous film, consisting of two different
types of conducting regions, denoted A and B, with areal fractions
$p_A$ and $p_B = 1-p_A$. In each region, the conductivity tensor is
that of a Drude metal in a transverse magnetic field, but the
density and the {\em sign} of charge carriers
can be different in the two regions.  We will show that, when $p_A =
1/2$, and the charge carriers have opposite signs the TMR is
asymptotically {\em exactly linear} at sufficiently strong magnetic
fields. Moreover, the linearity can extend down to quite low
magnetic fields.   The corresponding Hall coefficient $R_{H,e}$ is
found to vanish.  If $p_A \neq 1/2$, the effective resistivity
tensor ${\bf {\rho}_e }$ cannot be calculated exactly. An effective
medium approximation (EMA), which agrees with the exact result at
$p_A = 1/2$, predicts that the resistivity saturates for any $p_A
\neq 1/2$, and that $R_{H,e}$ changes sign at $p_A = 1/2$.   All
these results are in rough agreement with recent
experiments\cite{lee} [which are, however, carried out for
three-dimensional (3D) samples; see below].  If the carriers have
the {\em same} sign, no exact statements are possible, even at $p_A
= 1/2$.   But even in this case the EMA predicts a linear TMR
precisely at $p_A = 1/2$, though smaller than for carriers of
opposite sign.

We first prove the exact linearity of the TMR at $ p_A = 1/2$ for
carriers of opposite sign and opposite mobility, using a duality
argument.  We consider a two-dimensional (2D) conductor with a
spatially varying conductivity tensor ${\bf {\sigma}(x) }$, and
denote the effective conductivity tensor by ${\bf \sigma_e}$.
$\sigma_e$ is a $2 \times 2$ tensor defined by $\langle {\bf
J}\rangle = \sigma_e\langle{\bf E}\rangle$, where ${\bf J}$ and
${\bf E}$ are the position-dependent current density and electric
field, and $\langle...\rangle$ denotes a spatial average in the
limit of a large sample and suitable boundary conditions (as
discussed, for example, in Ref.\ \cite{stroud76}). $\sigma_e$ is the
quantity which would be measured as the sample conductivity in an
experiment. To calculate $\sigma_e$, we use a duality
theorem\cite{mendel}, which states that
\begin{equation}
\sigma_e[\sigma({\bf x})]
\sigma_e[\sigma^{-1}({\bf x})] = I,
\label{eq:dual}
\end{equation}
where $I$ is the 2$\times$2 unit matrix.  Here $\sigma_e[\sigma({\bf
x})]$ denotes the effective conductivity tensor of a material whose
local conductivity tensor is position-dependent and equal to
$\sigma({\bf x})$.

Thus, the product of $\sigma_e$ for the system of interest, and that
of a hypothetical ``dual composite'' whose local conductivity tensor
${\bf \sigma_d({\bf x})}$ is the local resistivity tensor of the
original material, equals the unit tensor.

We now apply this theorem to the following special case. Let the two
components each have a free-electron conductivity, but
carriers of opposite signs. For the first component 
\begin{equation}
\sigma_{A,xx}=\sigma_{A,yy} = \frac{\sigma_{A,0}}{1+H^2},
\label{eq:sigaxx}
\end{equation}
\begin{equation}
\sigma_{A,xy}=-\sigma_{A,yx} = \frac{\sigma_{A,0} H}{1+H^2},
\label{eq:sigaxy}
\end{equation}
where $\sigma_{A,0}$ is the zero-field conductivity. the
dimensionless magnetic field $H = \mu_AB/c$, where $\mu_A =
e\tau_A/m_A$ is an effective mobility of carriers of type A, $m_A$
is their effective mass, $e
> 0$ is the electron charge magnitude, and $\tau_A$ a
characteristic relaxation time. For the second component, we assume
\begin{equation}
\sigma_{B,xx} = \sigma_{B,yy} = \frac{\sigma_{B,0}}{1 + k^2H^2}.
\label{eq:sigbxx}
\end{equation}
\begin{equation}
\sigma_{B,xy} = -\sigma_{B,yx} = \frac{\sigma_{B,0}kH}{1+k^2H^2},
\label{eq:sigbxy}
\end{equation}
with the dimensionless constant $k = -1$ (i.\ e., the two types of
charge carriers have opposite signs).
We also introduce $\mu_B =k\mu_A$  as the effective mobility of
type-B carriers.   Finally, we assume that the composite contains an
areal fraction $p_i =1/2$ (i = A or B) of each component, and that
the geometry is symmetric. "Symmetric" means that, if the components
A and B were interchanged, $\sigma_e$ of the film will remain the
same in the thermodynamic limit.   There are many geometries, both
ordered (e. g. checkerboard) and random, which are symmetric by this
definition. If we make the usual Drude assumption that $\sigma_{i,0}
= n_ie|\mu_i|$ (i = A, B), where $n_i$ is the number density of
carriers of type $i$, then eqs.\ (\ref{eq:sigaxx})-(\ref{eq:sigbxy})
imply (i) that there are equal {\em areal fractions} of positive and
negative charge carriers (but not that the total numbers of positive
and negative charge carriers are equal); and (ii) that the
mobilities $\mu_A$ and $\mu_B$ are equal and opposite, so that
$\langle \mu \rangle = \sum_{i = A, B}p_i\mu_i = 0$.

Given these assumptions, the tensors $\sigma_A$ and $\sigma_B$
satisfy the remarkable relationship
\begin{equation}
\sigma_A^{-1} = \frac{1+H^2}{\sigma_0^2}\sigma_B,
\end{equation}
where $\sigma_0 = \left(\sigma_{A,0}\sigma_{B,0}\right)^{1/2}$.
Since we have an equal proportion of components A and B, distributed
in some symmetrical (and isotropic) fashion, the dual composite has
a conductivity tensor
\begin{equation}
\sigma_d({\bf x}) = \frac{1+H^2}{\sigma_0^2}\tilde{\sigma}({\bf x}),
\end{equation}
where $\tilde{\sigma}({\bf x})$ means the conductivity of a
composite in which the $A$ and $B$ components are interchanged.
Since $\sigma_d$ is just a multiple of the original conductivity
tensor $\sigma({\bf x})$, but with $A$ and $B$ components
interchanged, and since by the assumption of a symmetric composite
$\sigma_e[\sigma({\bf x})]=\sigma_e[\tilde{\sigma}({\bf x})]$, it
follows that
\begin{equation}
\sigma_e[\sigma_d({\bf x})] = \frac{1+H^2}{\sigma_0^2}
\sigma_e[\sigma({\bf x})].
\end{equation}

We now apply eq.\ (1) to this model, with the result
\begin{equation}
\frac{1 + H^2}{\sigma_0^2}\sigma_e^2[\sigma({\bf x})] = I.
\label{eq:sige}
\end{equation}
A physically acceptable solution to eq.\ (\ref{eq:sige}) must have
the diagonal elements of $\sigma_e$ equal and positive, and
off-diagonal elements equal and opposite.   It is readily shown
algebraically that the only such solution is
\begin{equation}
\sigma_e[\sigma({\bf x})] = \frac{1}{\sqrt{1+H^2}}\sigma_0 I.
\label{eq:sigd}
\end{equation}
The corresponding resistivity tensor $\rho_e$ is
\begin{equation}
\rho_e = \sigma_0^{-1}\sqrt{1+H^2}I.
\label{eq:rhod}
\end{equation}
The TMR is defined by the relation $\Delta\rho_{e,xx}(H) =
[\rho_{e,xx}(H)- \rho_{e,xx}(0)]/\rho_{e,xx}(0)$.  For this model,
$\Delta\rho_{e,xx}(H) = \sqrt{1+H^2}-1$ becomes linear in H for
large enough H, and the corresponding Hall coefficient $R_H =
\rho_{xy}(H)/H = 0$. Thus, this calculation appears to reproduce the
numerical results of Ref.\ \cite{parish}, but {\em analytically}.

Since the duality argument is not sufficient to determine $\sigma_e$
for $p_A \neq 1/2$, we have used the EMA for such concentrations.
The EMA is a simple mean-field approximation in which the local
electric fields and currents are calculated as if a given region is
surrounded by a suitably averaged environment. For the present model
the EMA becomes \cite{stroud75}
\begin{equation}
\sum_{i = A, B} p_i\delta\sigma_i(I - \Gamma\delta\sigma_i)^{-1} = 0.
\label{eq:ema}
\end{equation}
Here $\delta\sigma_i = \sigma_i - \sigma_e$, and $\Gamma$ is a
suitable depolarization tensor.  We assume that $\sigma_A$ and
$\sigma_B$ satisfy $\sigma_{i,xx} = \sigma_{i,yy}$; $\sigma_{i,xy} =
-\sigma_{i,yx}$.  Then the components of ${\bf \sigma_e}$ satisfy
$\sigma_{e,xx} = \sigma_{e,yy}$, $\sigma_{e,xy} = -\sigma_{e,yx}$.
We also assume that the two components $A$ and $B$ are distributed
in compact, approximately circular regions. Then $\Gamma =
-I/(2\sigma_{e,xx})$ \cite{stroud75}.   With these assumptions,
eqs.\ (\ref{eq:ema}) reduce to two coupled algebraic equations for
$\sigma_{e,xx}$ and $\sigma_{e,xy}$ which are easily solved
numerically.

To confirm that the EMA gives reasonable results, we have tested it
for $p_A = p_B = \frac{1}{2}$, and $\sigma_A$ and $\sigma_B$ given
by eqs.\ (\ref{eq:sigaxx})-(\ref{eq:sigbxy}) with $k = -1$.  We find
that the solution to the eq.\ (\ref{eq:ema}) for the tensor
$\sigma_e$ is diagonal, and a multiple of the unit tensor; the
diagonal elements are given by eq.\ (\ref{eq:sigd}). Thus, for $p_A
= p_B$, the EMA agrees with the exact duality arguments.

\begin{figure}
\includegraphics[width=0.8\columnwidth]{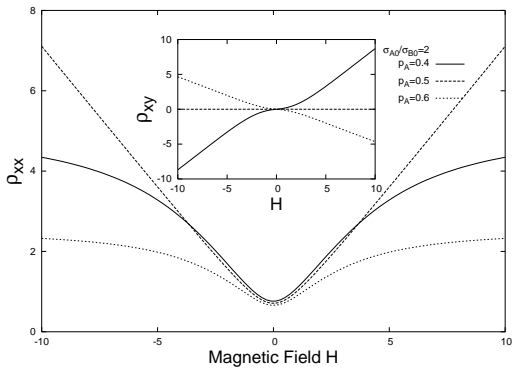}
\caption{Calculated
transverse resistivity $\rho_{e,xx}(H, p_A)$ and Hall resistivity
$\rho_{e,xy}(H, p_A)$ (inset) for a two-dimensional model
inhomogeneous semiconductor in a transverse magnetic field, as
calculated within the EMA for three different area fractions $p_A$
of component A.  Both are given in units of $1/\sigma_0 \equiv
1/\sqrt{\sigma_{A,0}\sigma_{B,0}}$.  The two components $A$ and $B$
have conductivities given by eqs.\ (2)-(5), with
$\sigma_{A,0}/\sigma_{B,0} =2$.  The mobilities of the two carriers
are assumed to have the same magnitudes: $|\mu_A| = |\mu_B|$.}
\end{figure}

To illustrate the EMA predictions for $p_A \neq \frac{1}{2}$, we
calculate $\sigma_e$ for $\sigma_i$ given by eqs.\
(\ref{eq:sigaxx})-(\ref{eq:sigbxy}). The resulting elements of the
resistivity tensor. $\rho_{e,xx} = \sigma_{e,xx}/[\sigma_{e,xx}^2 +
\sigma_{e,xy}^2]$, $\rho_{e,xy} = -
\sigma_{e,xy}/[\sigma_{e,xx}^2+\sigma_{e,xy}^2]$, are plotted in
Fig.\ 1 for $\sigma_{A0}/\sigma_{B0} = 2$. Evidently, and as can be
shown explicitly from the EMA equations, $\rho_{e,xx}$ is strictly
linear in $H$ only at $p_A = 1/2$. At all other concentrations,
$\rho_{e,xx}(H)$ saturates (i. e. approaches a constant) at large
$H$, but at a value much larger than $\rho_{e,xx}(H=0)$.   It is
easily shown that the saturation value of $\Delta\rho_{e,xx}(p_A)
\equiv$ Lim$_{H\rightarrow\infty}$
$[\rho_{e,xx}(H,p_A)/\rho_e(0,p_A) - 1 ]\propto 1/|p_A - p_c|$  on
both sides of the percolation threshold $p_c = 1/2$. Fig.\ 1 also
shows that the effective Hall resistivity $\rho_{e,xy}$ changes sign
just at the concentration where $\rho_{e,xx}$ varies asymptotically
linearly with $H$.

We have also solved the EMA for a composite described by eqs.\
(\ref{eq:sigaxx}) - (\ref{eq:sigbxy}) but for the more general case
in which $k \neq -1$.  Then $k>0$ and $k<0$ correspond respectively
to carriers with mobilities of the same and opposite signs.

\begin{figure}
\includegraphics[width=0.8\columnwidth]{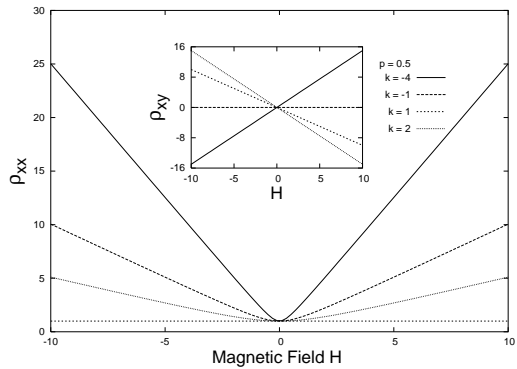}
\caption{ Same as
Fig.\ 1, but for $p_A = 1/2$, and various choices of the mobility
ratio $k = \mu_B/\mu_A$.  A positive or negative $k$ means that the
carriers have the same or opposite signs.}
\end{figure}

In Fig.\ 2 we show the EMA results for this model. Specifically,we
show $\rho_{xx}(H, p_A)$ and $\rho_{xy}(H,p_A)$ with $p_A = 1/2$,
$\sigma_{A,0} =\sigma_{B,0}$, and several choices of $k$
corresponding to carriers of both opposite and the same sign. The
case $k = 1$ actually corresponds to a {\em homogeneous} free
electron metal.  For all other values of $k$, the TMR is
asymptotically {\em linear}; the linear behavior is evident even at
moderate fields ($H \sim 1$). However, the linear {\em slope} is
larger when the carriers have opposite signs. We emphasize that
these results are obtained in the EMA. The duality arguments do not
give any predictions for $\rho_{xx}$ except when the carriers have
opposite signs and opposite mobilities.

In Figs.\ 3 and 4, we plot the resistivity $\rho_{xx}$ and Hall
coefficient $R_H \equiv \rho_{xy}/H$ as a function of $p_A$ for $H =
1$ and $H = 10$. In both cases, we assume that $\sigma_{A,0} =
\sigma_{B,0}$ and $|\mu_A| = |\mu_B|$.  $\rho_{xx}$ has a peak at
$p_A=1/2 $, which sharpens, as a function of $p_A$, as $H$
increases. Similarly, the Hall coefficient $R_H$ changes sign at
$p_A = 1/2$, and the change occurs over a narrower and narrower
regime of $p_A$ as $H$ increases.

\begin{figure}
\includegraphics[width=0.8\columnwidth]{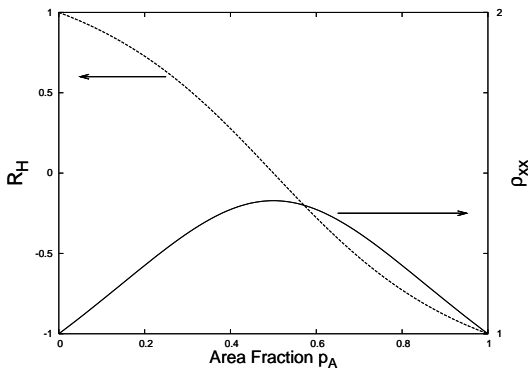}
\caption{Transverse resistivity $\rho_{xx}(H,p_A)$ and Hall
coefficient $R_H(H,p_A)$ as a function of $p_A$ for H = 1, using the
same model as in Fig. 1, with $\sigma_{B,0} = \sigma_{A,0}$ and
$\mu_B = -\mu_A$.}
\end{figure}

\begin{figure}
\includegraphics[width=0.8\columnwidth]{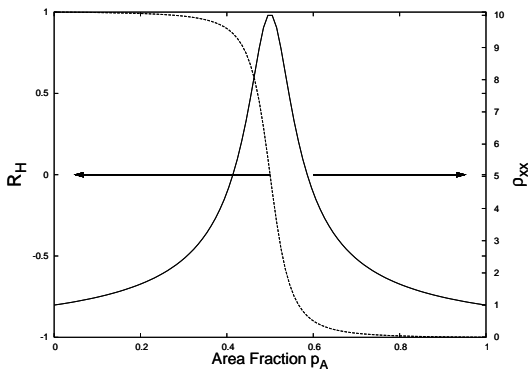}
\caption{ Same as
Fig.\ 3, but for $H = 10$.}
\end{figure}

The present results agree qualitatively with the experiments of Lee
{\it et al}\cite{lee}, which also show that the TMR peaks at
pressures where the Hall coefficient changes sign. But this
agreement should be viewed cautiously. In particular, the
measurements of Ref.\ \cite{lee} are carried out on a 3D sample,
while our calculations are for a 2D system.  The present work would
also apply to a 3D system with a columnar microstructure - that is,
a system in which the conductivity tensor $\sigma({\bf x})$ is
independent of the third dimension, $z$ - and the applied field
${\bf B} \| z$, but the samples of Ref.\ \cite{lee}, if
inhomogeneous, are most likely composed of small compact grains.  We
have calculated $\sigma_e$ for a 3D granular sample with carriers of
opposite signs, using the EMA, and find results similar to those
shown here for 2D samples.  These 3D calculations will be presented
elsewhere\cite{guttal1}.



The TMR of the present model is very large - $\Delta\rho_{xx}(H,
1/2) \sim 10$ for $H \sim 10$ - and remains approximately linear
down to fields as low as $H \sim 1-2$. By contrast, other models of
TMR which arises from inhomogeneities produce only a small TMR, or,
if a large TMR, $\Delta\rho_{xx}(H)$ does not vary linearly with
$H$\cite{stroud76,stroud79}.



In summary, we have presented a simple model of a 2D macroscopically
inhomogeneous material, whose TMR is asymptotically linear in
magnetic field, and whose corresponding Hall coefficient vanishes.
The model has several unusual properties which make it likely to be
realized only in special circumstances. First, eqs.\
(\ref{eq:sigaxx})-(\ref{eq:sigbxy})) imply that the carriers have
equal and opposite mobilities $\mu_A = -\mu_B$. Secondly, the
linearity occurs only if the composite has a symmetric geometry at
$p_A = 1/2$. But given these features,the TMR, arising from a
perpendicular to the sample, is asymptotically {\em exactly} linear
in B. To our knowledge, this is the only analytically soluble model
for TMR due to macroscopic inhomogeneities, which produces a linear
TMR at high concentrations of inhomogeneities.

This work was supported by the National Science Foundation, through
grant DMR04-13395.  We also benefited from the facilities of the
Ohio Supercomputer Center.






\end{document}